\RequirePackage{ifpdf}
\ifpdf 
\documentclass[pdftex]{sigma}
\else
\documentclass{sigma}
\fi

\numberwithin{equation}{section}

\begin{document}

\allowdisplaybreaks

\renewcommand{\thefootnote}{$\star$}

\renewcommand{\PaperNumber}{002}

\FirstPageHeading

\ShortArticleName{On a Nonlocal Ostrovsky--Whitham Type Dynamical System}

\ArticleName{On a Nonlocal Ostrovsky--Whitham Type\\ Dynamical System, Its Riemann
Type Inhomogeneous\\ Regularizations and Their Integrability\footnote{This
paper is a contribution to the Proceedings of the Eighth
International Conference ``Symmetry in Nonlinear Mathematical
Physics'' (June 21--27, 2009, Kyiv, Ukraine). The full collection
is available at
\href{http://www.emis.de/journals/SIGMA/symmetry2009.html}{http://www.emis.de/journals/SIGMA/symmetry2009.html}}}

\Author{Jo\l anta GOLENIA~$^{\dag^1}$, Maxim V. PAVLOV~$^{\dag^2}$,\\ Ziemowit POPOWICZ~$^{\dag^3}$ and Anatoliy K. PRYKARPATSKY~$^{\dag^4\dag^5}$}

\AuthorNameForHeading{J.~Golenia, M.V.~Pavlov, Z.~Popowicz and A.K.~Prykarpatsky}

\Address{$^{\dag^1}$~The Department  of Applied Mathematics,  AGH
University of Science and Technology,\\
\hphantom{$^{\dag^1}$}~Krak\'ow 30059, Poland}
\EmailDD{\href{mailto:golenia@agh.edu.pl}{golenia@agh.edu.pl}}

\Address{$^{\dag^2}$~Department of Mathematical Physics, P.N. Lebedev Physical Institute,\\
\hphantom{$^{\dag^2}$}~53 Leninskij Prospekt,  Moscow 119991, Russia}
\EmailDD{\href{mailto:M.V.Pavlov@lboro.ac.uk}{M.V.Pavlov@lboro.ac.uk}}

\Address{$^{\dag^3}$~The Institute for Theoretical Physics,   University of Wroc\l aw,   Wroc\l aw 50204, Poland}
\EmailDD{\href{mailto:pryk.anat@ua.fm}{ziemek@ift.uni.wroc.pl}}
\URLaddressDD{\url{www.ift.uni.wroc.pl/~ziemek/}}

\Address{$^{\dag^4}$~The Department of Mining Geodesics,   AGH University
of Science and Technology,\\
\hphantom{$^{\dag^4}$}~Krak\'ow 30059, Poland}

\Address{$^{\dag^5}$~Department of Economical Cybernetics, Ivan Franko State Pedagogical University,\\
\hphantom{$^{\dag^5}$}~Drohobych, Lviv Region, Ukraine}
\EmailDD{\href{mailto:pryk.anat@ua.fm}{pryk.anat@ua.fm}}

\ArticleDates{Received October 14, 2009, in f\/inal form January 03, 2010;  Published online January 07, 2010}

\Abstract{Short-wave perturbations in a relaxing medium, governed by a special
reduction of the Ostrovsky evolution equation, and later derived by Whitham,
are studied using the gradient-holonomic integrability algorithm. The
bi-Hamiltonicity and complete integrability of the corresponding dynamical
system is stated and an inf\/inite hierarchy of commuting to each other
conservation laws of dispersive type are found. The well def\/ined
regularization of the model is constructed and its Lax type integrability is
discussed. A generalized hyd\-rodynamical Riemann type system is considered,
inf\/inite hierarchies of conservation laws, related compatible Poisson
structures and a Lax type representation for the special case $N=3$ are
constructed.}


\Classification{35C05; 37K10}

\section{Introduction}

Many important problems of propagating waves in nonlinear media with
distributed parameters, for instance, invisible non-dissipative dark matter,
playing a decisive role \cite{GZ,GZ1} in the formation of large scale
structure in the Universe like galaxies, clusters of galaxies,
super-clusters, can be described by means of evolution dif\/ferential
equations of special type. It is also well known \cite{BPrGG,MPV,PrPryt,Wh}
that shortwave perturbations in a relaxing one dimensional medium can be
described by means of some reduction of the Ostrovsky equations, coinciding
with the Whitham type evolution equation
\begin{gather}
du/dt=2uu_{x}+\int_{\mathbb{R}}\mathcal{K}(x,s)u_{s}ds,  \label{W1.1}
\end{gather}
discussed f\/irst in \cite{Wh}. Here the kernel $\mathcal{K}:\mathbb{R}\times
\mathbb{R}\rightarrow \mathbb{R}$ depends on the medium elasticity
properties with spatial memory and can, in general, be a function of the
pressure gradient $u_{x}\in C^{\infty }(\mathbb{R};\mathbb{R}),$ evolving
with respect to equation (\ref{W1.1}). In particular, if the nonlinear
medium is endowed still with spatial memory properties, that is the wave
amplitude depends on the orbit, swept by its front, the propagation of the
corresponding wave can be modeled by means of the so called
generalized Ostrovsky evolution equations \cite{Os}. Namely, if to put $%
\mathcal{K}(x,s)=\frac{1}{2}|x-s|$, $x,s\in \mathbb{R},$ then equation~(\ref{W1.1}) can be reduced to
\begin{gather*}
du/dt=2uu_{x}+\partial ^{-1}u, 
\end{gather*}
which was, in particular, studied before in \cite{MPV,Pa,Pa1}.

Since some media possess elasticity properties depending strongly on the
spatial pressure gradient $u_{x},~x\in \mathbb{R},$ the corresponding
Whitham type kernel looks like
\begin{gather}
\mathcal{K}(x,s):=-\theta (x-s)u_{s}  \label{W1.3}
\end{gather}
for $x,s\in \mathbb{R},$ naturally modeling the relaxing spatial memory
ef\/fects. The resulting equation~(\ref{W1.1}) with the kernel (\ref{W1.3})
becomes
\begin{gather}
du/dt=2uu_{x}-\partial ^{-1}u_{x}^{2}:=K[u],  \label{W1.4}
\end{gather}
which appears to possess very interesting mathematical properties. The
latter will be the main topic of the next sections below.

Owing to the results, obtained before in \cite{PrPryt,PP}, the dynamical
system (\ref{W1.4}) appeared to be a~Lax type integrable bi-Hamiltonian
f\/low, but with ill posed temporal evolution. As it was demonstrated in \cite{PP}, a suitable f\/inite-dimensional reduction scheme, if applied to the
correspon\-ding hierarchy of conservation laws for constructing explicit
solutions to the Ostrovsky--Whitham type nonlinear dynamical system (\ref{W1.4}) by means of quadratures, meets some technical problems. Some of
these integrability aspects were before presented in \cite{BPrGG}, where a
suitable well posed regularization of the equation (\ref{W1.4}) in the form
\begin{gather}
\left.
\begin{array}{@{}l}
u_{t}=2uu_{x}-v \\
v_{t}=2uv_{x}
\end{array}
\right\} :=K[u,v]  \label{W1.4a}
\end{gather}
for treating this nonlocality problem was proposed.

Below the well posed integrability problem for the Ostrovsky--Whitham type
nonlinear and nonlocal dynamical system (\ref{W1.4a}) will be reanalyzed in
detail making use of this regularization scheme. The corresponding implectic
structures and Lax type representations are found by means of the
dif\/ferential-geometric tools, devised and extended in \cite{FT,FF,Ma,PM}. A
natural Riemann type generalization of the dynamical system (\ref{W1.4a}) is
proposed, owing to a recent observation by D.~Holm and M.~Pavlov:
\begin{gather}
D_{t}^{N}u=0,\qquad N\in \mathbb{Z}_{+},  \label{W1.4b}
\end{gather}%
which at $N=2$ is exactly equivalent to the system (\ref{W1.4a}). The
integrability properties of equation~(\ref{W1.4b}) at $N=3$ were analyzed in
detail, the conservation laws, corresponding compatible implectic structures
and Lax type representation are constructed.

It is worth to mention that the obtained in this work Lax type pair~(\ref{W5.3}) for the regularized dynamical system~(\ref{W1.4a}) was found f\/irst
in work \cite{DB}. It coincides with those found later in~\cite{Pav}, making
use of a very special bi-Lagrangian representation of the dynamical system~(\ref{W1.4a}). But the existence of the singular co-implectic structure (\ref{W5.8-b}) in these references was not stated. A~detailed analysis of the
relationships between solutions of dynamical systems (\ref{W1.4}) and (\ref{W1.4a}), based on a reciprocal transformation, suggested by M.~Pavlov in~\cite{Pav}, was presented recently in~\cite{Sak}. Mention also work~\cite{Le}, where the geometric aspects of the equation like~(\ref{W5.2}) were
studied.

Note also here that theory of integrable homogenous hydrodynamic type
systems with distinct characteristic velocities was constructed by S.P.~Tsarev. In this paper we consider the f\/irst example in a literature of
nonhomogeneous integrable hydrodynamic type systems with a sole
characteristic velocity. Such a theory does not exist at this moment.

\section{A regularization scheme and the geometric\\ integrability problem}

Def\/ine a smooth periodic function $v\in C_{2\pi }^{\infty }(\mathbb{R};\mathbb{R}),$ such that%
\begin{gather*}
v:=\partial ^{-1}u_{x}^{2}  
\end{gather*}%
for any $x,t\in \mathbb{R},$ where the function $u\in C_{2\pi
}^{\infty}(\mathbb{R};\mathbb{R})$ solves equation~(\ref{W1.4}).
Then it is easy to state that the following regularized nonlinear dynamical
system
\begin{gather}
\left.
\begin{array}{@{}l}
u_{t}=2uu_{x}-v \\
v_{t}=2uv_{x}
\end{array}
\right\} :=K[u,v]  \label{W5.2}
\end{gather}
of hydrodynamic type, which was introduced before in \cite{Dav}, studied in
\cite{BPrGG,DB,HS,HZ,OR} and analyzed as a Gurevich--Zybin system in \cite{Pav}, and is already well def\/ined on the extended $2\pi $-periodic
functional space $\mathcal{M}:=C_{2\pi }^{\infty }(\mathbb{R};\mathbb{R}%
^{2}) $ and equivalent on the functional submanifold $\mathcal{M}_{\rm red}:=\{(u,v)\in \mathcal{M}:v_{x}-u_{x}^{2}=0\}$ to that given by
expression (\ref{W1.4}), as it was mentioned in \cite{BPrGG} and discussed
recently in \cite{Sak}. The system (\ref{W5.2}) can be rewritten as the
following set of equations
\begin{gather}
u_{t}=2uu_{x}-v,\qquad v_{t}=2uv_{x},\nonumber \\
u_{x}=w,\text{ }v_{x}=u_{x}w,\nonumber \\
w_{t}=v_{x}+2uw_{x},
  \label{W5.2a}
\end{gather}
which is equivalent to a set of dif\/ferential two-forms
\begin{gather}
\{\alpha \}:=\big\{\alpha ^{(1)}=du\wedge dx+2u du\wedge dt-v dx\wedge
dt,\  \alpha ^{(2)}=dv\wedge dx+2u dv\wedge dt,\nonumber
\\
\phantom{\{\alpha \}:=\{}{} \alpha ^{(3)}=du\wedge dt-w dx\wedge dt,\  \alpha ^{(4)}=dv\wedge
dt-w du\wedge dt,\nonumber\\
\phantom{\{\alpha \}:=\{}{} \alpha ^{(5)}=dw\wedge dx+dv\wedge dt+2u dw\wedge dt\big\}. \label{5.2aa}
\end{gather}
This set of two-forms generates the closed ideal $\mathcal{I}(\alpha ),$
since
\begin{gather*}
d\;\alpha ^{(1)}=-\alpha ^{(2)}\wedge dt,\qquad d \alpha ^{(2)}=2du\wedge
\alpha ^{(4)},\qquad d \alpha ^{(3)}=-\alpha ^{(5)}\wedge dt,\\
d\alpha ^{(4)}=-dw\wedge \alpha ^{(3)}-wdt\wedge \alpha ^{(5)},\qquad
d \alpha ^{(5)}=-2dw\wedge \alpha ^{(3)}-2w dt\wedge \alpha ^{(5)}.
\end{gather*}

The set of dif\/ferential forms (\ref{5.2aa}), being integrable, def\/ines the
integral submanifold~$\bar{M}$   by means of the condition $\mathcal{I}(\alpha )=0.$
 Making now use of the dif\/ferential-geometric method devised in~\cite{HPP,Ol,PM} and extending algorithmically the approach of~\cite{Ma}, we
will look for a reduced upon the integral submanifold $\bar{M}$ connection
one-form $\Gamma ,$ belonging to some not yet determined its holonomy Lie
algebra $\mathcal{G}.$ This 1-form can be represented as follows:
\begin{gather}
\Gamma ={\mathcal{A}}(u,v,w) dx+{\mathcal{B}}(u,v,w) dt,  \label{W5.5b}
\end{gather}%
where the elements $\mathcal{A}, \mathcal{B}\in \mathcal{G}$ satisfy
determining equations
\begin{gather}
\Omega =\frac{\partial \mathcal{A}}{\partial u} du\wedge dx+\frac{\partial
\mathcal{A}}{\partial v} dv\wedge dx+\frac{\partial \mathcal{A}}{\partial w}%
 dw\wedge dx+\frac{\partial \mathcal{B}}{\partial u} du\wedge dt\nonumber\\
\phantom{\Omega =}{}
+\frac{\partial \mathcal{B}}{\partial v}\;dv\wedge dt+\frac{\partial
\mathcal{B}}{\partial w}\;dw\wedge dt+[\mathcal{A},\mathcal{B}]dx\wedge
dt\nonumber\\
\Rightarrow \quad g_{1}(du\wedge dx+2u du\wedge dt-v dx\wedge
dt) + g_{2}(dv\wedge dx+2u dv\wedge dt) \nonumber\\
\qquad{} +g_{3}(du\wedge dt-w dx\wedge dt) + g_{4}(dv\wedge dt-w du\wedge
dt) \nonumber\\
\qquad{} +g_{5}(dw\wedge dx+2u dw\wedge dt+dv\wedge dt)\in \mathcal{I}(\alpha
)\otimes \mathcal{G}\label{W5.5a}
\end{gather}
for some $\mathcal{G}$-valued functions $g_{1},\dots,g_{5}\in \mathcal{G}$ on~$M.$ From (\ref{W5.5a}) one f\/inds that
\begin{gather}
\frac{\partial \mathcal{A}}{\partial u}=g_{1},\qquad \frac{\partial \mathcal{A}}{\partial v}=g_{2},
\qquad \frac{\partial \mathcal{A}}{\partial w}%
=g_{5},\nonumber
\\
\frac{\partial \mathcal{B}}{\partial u}=2u g_{1}+g_{3}-w g_{4},\qquad \frac{\partial \mathcal{B}}{\partial v}=2u g_{2}+g_{4}+g_{5},
\nonumber\\
\frac{\partial \mathcal{B}}{\partial w}=2u g_{5},\qquad \lbrack \mathcal{A},\mathcal{B}]=-vg_{1}-wg_{3}.\label{W5.6a}
\end{gather}
Thereby, from the obtained set of relationships (\ref{W5.6a}) one can f\/ind
that
\begin{gather*}
\mathcal{B}=2u\mathcal{A}+\mathcal{C}(u,v),\qquad g_{4}=\frac{\partial
\mathcal{C}}{\partial v}-\frac{\partial \mathcal{A}}{\partial w},\qquad
g_{3}=2\mathcal{A}+\frac{\partial \mathcal{C}}{\partial u}+w\frac{\partial
\mathcal{C}}{\partial v}-w\frac{\partial \mathcal{A}}{\partial w},
\nonumber\\
\lbrack \mathcal{A},\mathcal{C}]=-v\frac{\partial \mathcal{A}}{\partial u}-2w
\mathcal{A}-w\frac{\partial \mathcal{C}}{\partial u}-w^{2}\frac{\partial
\mathcal{C}}{\partial v}+w^{2}\frac{\partial \mathcal{A}}{\partial w},
\end{gather*}
serving for f\/inal searching for connection (\ref{W5.5b}).

\section{The bi-Hamiltonian structure and Lax-type representation}

Consider the following polynomial expansion of the element $\mathcal{A}(u,v;w)\in \mathcal{G}$ with respect to the variable $w$:
\begin{gather*}
\mathcal{A}=\mathcal{A}_{0}(u,v)+\mathcal{A}_{1}(u,v)w+\mathcal{A}
_{2}(u,v)w^{2}  
\end{gather*}
and substitute it into the last equation of (\ref{W5.6a}). As a result we
obtain:
\begin{gather}
\lbrack \mathcal{A}_{0},C]  = -v\frac{\partial \mathcal{A}_{0}}{\partial u},\qquad [\mathcal{A}_{1},C]=-v\frac{\partial \mathcal{A}_{1}}{\partial u}-2\mathcal{A}_{0}-\frac{\partial C}{\partial u}, \nonumber  \\
\lbrack \mathcal{A}_{2},C] =-v\frac{\partial \mathcal{A}_{2}}{\partial u}-
\frac{\partial C}{\partial v}-\mathcal{A}_{1},  \label{W5.8a}
\end{gather}%
or
\begin{gather}
\mathcal{A}_{1}=[C,\mathcal{A}_{2}]-v\frac{\partial \mathcal{A}_{2}}{
\partial u}-\frac{\partial C}{\partial v}.  \label{W5.9a}
\end{gather}%
which can be substituted into the second equation of (\ref{W5.8a}):
\begin{gather*}
\lbrack \lbrack C,\mathcal{A}_{2}],C]-2v\left[\frac{\partial \mathcal{A}_{2}}{\partial u},C\right]-\left[\frac{\partial C}{\partial v},C\right]
 =-v\left[\frac{\partial C}{\partial u},\mathcal{A}_{2}\right]-v^{2}\frac{\partial ^{2}\mathcal{A}_{2}}{\partial u^{2}}-v\frac{\partial
^{2}C}{\partial u\partial v}-2\mathcal{A}_{0}-\frac{\partial C}{\partial u}.
\end{gather*}
Thus, recalling (\ref{W5.8a}) and (\ref{W5.9a}), we have that%
\begin{gather}
2\mathcal{A}_{0} =[C,[C,\mathcal{A}_{2}]]+2v\left[\frac{\partial \mathcal{A}_{2}%
}{\partial u},C\right]+   \left[\frac{\partial C}{\partial v},C\right]-v\left[\frac{\partial C}{\partial u},%
\mathcal{A}_{2}\right]-  v^{2}\frac{\partial ^{2}\mathcal{A}_{2}}{\partial u^{2}}-v\frac{\partial
^{2}C}{\partial u\partial v}-\frac{\partial C}{\partial u},  \nonumber \\
\lbrack \mathcal{A}_{0},C] =-v\frac{\partial \mathcal{A}_{0}}{\partial u},
\qquad \mathcal{A}_{1}=[C,\mathcal{A}_{2}]-v\frac{\partial \mathcal{A}
_{2}}{\partial u}-\frac{\partial C}{\partial v}. \label{W5.11a}
\end{gather}
Now we will assume that the element $C:=C_{0}$ is constant and the elements
$\mathcal{A}_{0}$ and $\mathcal{A}_{2}$ are linear with respect to variables $u$ and $v$, that is
\begin{gather*}
\mathcal{A}_{0} =\mathcal{A}_{0}^{(0)}+\mathcal{A}_{0}^{(1)}u+\mathcal{A}
_{0}^{(2)}v,   \qquad
\mathcal{A}_{2} =\mathcal{A}_{2}^{(0)}+\mathcal{A}_{2}^{(1)}u+\mathcal{A}
_{2}^{(2)}v.  
\end{gather*}
Whence and from (\ref{W5.11a}) one gets:
\begin{gather}
2\mathcal{A}_{^{0}}^{(0)}  = [C_{0},[C_{0},\mathcal{A}_{2}^{(0)}]], \qquad [\mathcal{A}_{0}^{(1)},C_{0}]=0, \qquad [\mathcal{A}_{0}^{(2)},C_{0}]=-\mathcal{A}_{0}^{(1)},
\nonumber \\
2\mathcal{A}_{^{0}}^{(1)}  = [C_{0},[C_{0},\mathcal{A}_{2}^{(1)}]],\qquad 2\mathcal{A}_{0}^{(2)}=[C_{0},[C_{0},\mathcal{A}_{2}^{(2)}]]+2[\mathcal{A}_{2}^{(1)},C_{0}].  \label{W5.13a}
\end{gather}
To solve the algebraic system (\ref{W5.13a}) we need to calculate \cite{PM}
the corresponding holonomy Lie algebra of the connection (\ref{W5.5b}). As a
result of simple, but slightly cumbersome calculations, we derive that
elements $\mathcal{A}_{2}^{(j)}$, $j=0,\dots,2$, and~$C_{0}$ belong to the
Lie algebra $sl(2;\mathbb{C})$, whose basis~$L_{0}$,~$L_{+}$ and~$L_{-}$ can be
taken to satisfy the following canonical commutation relations:
\begin{gather*}
\lbrack L_{0},L_{\pm }]=\pm L_{\pm },\qquad [L_{+},L_{-}]=2L_{0}.  
\end{gather*}%
Thereby, making use of the standard determining expansions%
\begin{gather}
\mathcal{A}_{2}^{(j)}  = \sum\limits_{\pm }c_{\pm }^{(j)}L_{\pm
}+c_{0}^{(j)}L_{0},   \qquad
C_{0}  = \sum\limits_{\pm }k_{\pm }L_{\pm }+k_{0}L_{0},  \label{W5.14aa}
\end{gather}
where $j=0,\dots,2,$ and substituting (\ref{W5.14aa}) into (\ref{W5.13a}), we obtain some relationships on values $c_{\pm }^{(j)},c_{0}^{(j)}\in
\mathbb{C}$, $j=0,\dots,2,$ and $k_{\pm },k_{0}\in \mathbb{C}.$
Resolving by means of simple but slightly cumbersome calculations these
relationships, we f\/ind the searched for basic elements $\mathcal{A}$ and $\mathcal{B}$ of the connection~$\Gamma ,$ depending on a spectral parameter $\lambda \in \mathbb{C},$ thereby giving rise to the correspon\-ding Lax type
commutative spectral representation for dynamical system~(\ref{W5.2}) in the
following $(2\times 2)$-matrix form:
\begin{gather}
\frac{df}{dx}=\ell \lbrack u,v;\lambda ]f,\qquad \frac{df}{dt}=p(\ell
)f,\qquad p(\ell ):=2u\ell \lbrack u,v;\lambda ]+q, \label{W5.3}\\
\ell \lbrack u,v;\lambda ]:=\left(
\begin{array}{cc}
-\lambda u_{x} & -v_{x} \\
\lambda ^{2} & \lambda u_{x}%
\end{array}
\right) ,\qquad q:=\left(
\begin{array}{cc}
0 & 0 \\
\lambda & 0%
\end{array}%
\right) , \qquad
p(\ell )=\left(
\begin{array}{cc}
-2\lambda u_{x}u & -2v_{x}u \\
\lambda +2\lambda ^{2}u & 2\lambda u_{x}u
\end{array}%
\right) ,
\nonumber
\end{gather}%
def\/ining the generalized time-independent spectrum ${\rm Spec}(\ell )\subset
\mathbb{C}$:   $\lambda \in {\rm Spec}(\ell )$,  if the correspon\-ding solution $%
f\in L_{\infty }(\mathbb{R};\mathbb{C}^{2}\mathbb{)}.$ It is worth to remark
here that the Lax type representation~(\ref{W5.3}), found for the dynamical
system~(\ref{W5.2}), is not unique. Moreover, making use of other imbeddings
of the connection form~(\ref{W5.5b}) into a suitable holonomy Lie algebra $\mathcal{G},$ one can construct dif\/ferent Lax type representations, which
could appear to be more useful for f\/inding exact solutions to dynamical
system~(\ref{W5.2}) by means of, for instance, the inverse spectral
transform method.

The standard Riccati equation, derived from (\ref{W5.3}), allows to obtain
right away an inf\/inite hierarchy of local conservation laws:%
\begin{gather}
\hat{\gamma}_{-1}:=\int_{0}^{2\pi }\sqrt{u_{x}^{2}-v_{x}}dx,\qquad \hat{\gamma}_{0}:=\int_{0}^{2\pi }\frac{(u_{x}v_{xx}-v_{x}u_{xx})}{2v_{x}\sqrt{u_{x}^{2}-v_{x}}}dx,\qquad \dots,  \label{W5.4}
\end{gather}
and so on. All of conservation laws (\ref{W5.4}) except~$\gamma _{-1},$ are
singular at the Cauchy condition~(\ref{W5.2a}). This means that we need to
construct other hierarchy of polynomial conservation laws regular on the
functional submanifold
\begin{gather}
\mathcal{M}_{\rm red}:=\big\{(u,v)\in \mathcal{M}:u_{x}^{2}-v_{x}=0,\  x\in
\mathbb{R}/2\pi \mathbb{Z}\big\}.  \label{W5.5}
\end{gather}%
The latter exists owing to the results of~\cite{Pav,PM}. The simplest way to
search for them consists in determining the bi-Hamiltonian structure of f\/low~(\ref{W5.2}). As it is easy to check, dynamical system~(\ref{W5.2}) is
canonically Hamiltonian, that is
\begin{gather*}
\frac{d}{dt}(u,v)^{\intercal }:=-\hat{\vartheta}\,{\rm grad}\,\hat{H}%
_{\vartheta }=\hat{K}[u,v],  
\end{gather*}%
where the corresponding co-symplectic structure $\hat{\vartheta}:T^{\ast }(\mathcal{M})\rightarrow T(\mathcal{M})$ is canonical, equals%
\begin{gather}
\hat{\vartheta}=\left(
\begin{array}{cc}
0 & 1 \\
-1 & 0
\end{array}
\right)  \label{W5.7}
\end{gather}%
and satisf\/ies the Noether equation
\begin{gather*}
L_{\hat{K}}\hat{\vartheta}=0=d\hat{\vartheta}/dt-\hat{\vartheta}\hat{K}
^{\prime ,\ast }-\hat{K}^{\prime }\hat{\vartheta}.  
\end{gather*}%
To prove this, it is enough to f\/ind by means of the small parameter method,
devised before in~\cite{PM} a non-symmetric $(\varphi ^{\prime }\neq \varphi
^{\prime ,\ast })$ solution $\varphi \in T(\mathcal{M})$ to the following
Lie--Lax equation:
\begin{gather}
d\varphi /dt+\hat{K}^{\prime ,\ast }\varphi ={\rm grad}\, L  \label{W5.7-1}
\end{gather}%
for some suitably chosen smooth functional $L\in \mathcal{D}(M).$ As a
result of easy calculations one obtains that
\begin{gather}
\varphi =(-v,0)^{\intercal },\qquad L=-\int_{0}^{2\pi }uvdx.
\label{W5.7-2}
\end{gather}%
Making use of (\ref{W5.7-2}) and the classical Legendrian relationship for
the suitable Hamiltonian function
\begin{gather}
H:=(\varphi ,\hat{K})-L,  \label{W5.7-4}
\end{gather}%
and the corresponding symplectic structure
\begin{gather}
\hat{\vartheta}^{-1}:=\varphi ^{\prime }-\varphi ^{\prime ,\ast }=\left(
\begin{array}{cc}
0 & -1 \\
1 & 0%
\end{array}
\right)  \label{W5.7-3}
\end{gather}
one obtains the implectic structure (\ref{W5.7}) and the corresponding
non-singular Hamilton function
\begin{gather*}
\hat{H}_{\vartheta }:=\int_{0}^{2\pi }(v^{2}/2+v_{x}u^{2})dx.  
\end{gather*}
It is here worth to mention that the determining Lie--Lax equation (\ref{W5.7-1}) possesses still another solution
\begin{gather*}
\varphi =\left(\frac{u_{x}}{2},-\frac{u_{x}^{2}}{2v_{x}}\right),\qquad L=\frac{1}{4}%
\int_{0}^{2\pi }uv_{x}dx,  
\end{gather*}%
giving rise, owing to formulas (\ref{W5.7-3}) and (\ref{W5.7-4}) to the new
co-implectic (singular ``symplectic'') structure
\begin{gather}
\hat{\eta}^{-1}:=\varphi ^{\prime }-\varphi ^{\prime ,\ast }=\left(
\begin{array}{cc}
\partial & -\partial u_{x}v_{x}^{-1} \vspace{2mm}\\
-u_{x}v_{x}^{-1}\partial & \frac{1}{2}(u_{x}^{2}v_{x}^{-2}\partial +\partial
u_{x}^{2}v_{x}^{-2})%
\end{array}%
\right)  \label{W5.8-b}
\end{gather}
and the Hamiltonian functional%
\[
\hat{H}_{\eta }:=\frac{1}{2}\int_{0}^{2\pi }(uv_{x}-vu_{x})dx.
\]
The co-implectic structure (\ref{W5.8-b}) is, evidently, singular since $%
\hat{\eta}^{-1}(u_{x},v_{x})^{\intercal }=0.$ Remark also that, owing to the
general symplectic theory results \cite{Bl,FF,HPP,Ma,MBPS,Ol,PM} for
nonlinear dynamical systems on smooth functional manifolds, operator (\ref{W5.8-b}) def\/ines on the manifold $\mathcal{M}$ a closed dif\/ferential
two-form. Thereby it is \textit{a priori} co-symplectic, satisfying on $%
\mathcal{M}$ the standard Jacobi brackets condition. Moreover, the implectic
structure $\hat{\eta}:T^{\ast }(\mathcal{M})\rightarrow T^{\ast }(\mathcal{M}%
)$ satisf\/ies the determining Noether equation
\begin{gather*}
L_{\hat{K}}\hat{\eta}=0=d\hat{\eta}/dt-\hat{\eta}\hat{K}^{\prime ,\ast }-\hat{K}^{\prime }\hat{\eta}, 
\end{gather*}
whose solutions can also be obtained by means of the small parameter method,
devised before in \cite{MBPS,PM}. As a result, the second implectic operator
has the form
\begin{gather}
\hat{\eta}:=\left(
\begin{array}{cc}
\partial ^{-1} & 2u_{x}\partial ^{-1} \\
2\partial ^{-1}u_{x} & 2v_{x}\partial ^{-1}+2\partial ^{-1}v_{x}%
\end{array}%
\right) ,  \label{W5.10}
\end{gather}%
giving rise to a new inf\/inite hierarchy of polynomial conservation laws
\begin{gather}
\hat{\gamma}_{n}:=\int_{0}^{1}d\lambda \langle (\hat{\vartheta}^{-1}\hat{\eta})^{n}{\rm grad}\, \hat{H}_{\vartheta }[u\lambda \},u\rangle  \label{W5.11}
\end{gather}
for all $n\in \mathbb{Z}_{+}.$

In particular, one can easily observe that there hold representations
\begin{gather*}
\frac{d}{dt}(u,v)^{\intercal }=-\hat{\eta}\,{\rm grad}\,\hat{H}%
_{\eta },\qquad \frac{d}{dx}(u,v)^{\intercal }=-\hat{\vartheta}\,  {\rm grad}\, \hat{H}_{\eta },  
\end{gather*}
where
\begin{gather*}
\hat{H}_{\eta }:=\frac{1}{2}\int_{0}^{2\pi }(uv_{x}-vu_{x})dx.  
\end{gather*}
Thereby, one can formulate the following proposition.

\begin{proposition} \label{Pr_1} The Riemann type hydrodynamical system \eqref{W5.2}
is a Lax type integrable bi-Hamiltonian flow on the functional manifold $\mathcal{M}$. The corresponding implectic pairs are compa\-tible and given by
matrix operators \eqref{W5.7} and \eqref{W5.10}, the Lax type representation
is presented in the differential matrix form~\eqref{W5.3}.
\end{proposition}

Now, making use of (\ref{W5.11}), one can apply the standard reduction
procedure upon the corresponding f\/inite dimensional functional subspaces $%
\mathcal{M}^{2n}\subset \mathcal{M}$, $n\in \mathbb{Z}_{+},$ and obtain a
large set of exact solutions of special quasi-periodic and solitonic type to
dynamical system (\ref{W5.2}) upon the functional submanifold $\mathcal{M}_{\rm red}$, if the Cauchy data are taken to satisfy constraint (\ref{W5.5}).
Here we need to mention that a general solution to the system (\ref{W5.2}),
obtained in \cite{Pav,Sak}, is presented in an unwieldy involved form,
almost completely not feasible for practical applications.

\section{A Riemann type hydrodynamical generalization}

It is here interesting to mention (owing to recent observations by D.~Holm
for $N=2$ and for arbitrary $N\in \mathbb{Z}_{+}$ by M.~Pavlov) that the dynamical
system (\ref{W5.2}) can be equivalently rewritten up to the time rescaling
as
\begin{gather}
D_{t}^{2}u=0,\qquad D_{t}:=\partial /\partial t+u\partial ,
\label{W5.14}
\end{gather}
under the f\/low velocity condition $dx/dt:=u$, which is a partial case \cite{MC} of the generalized Riemann type hydrodynamic system
\begin{gather}
D_{t}^{N}u=0  \label{W5.14b}
\end{gather}%
for any integer $N\in \mathbb{Z}_{+}.$ If $N=3,$ having def\/ined the new
variables $v:=D_{t}u,$ $z:=D_{t}v$, one easily obtains the new
dynamical system%
\begin{gather}
\left.
\begin{array}{@{}l}
u_{t}=v-uu_{x} \\
v_{t}=z-uv_{x} \\
z_{t}=-uz_{x}
\end{array}
\right\} :=K[u,v,z]  \label{W5.15}
\end{gather}%
of hydrodynamical type, which proves also to possess inf\/inite hierarchies of
polynomial conservation laws.

As we are interested f\/irst in the conservation laws for the system (\ref{W5.15}), the following proposition holds.
\begin{proposition}
\label{Pr_1a} Let $H(\lambda ):=\int_{0}^{2\pi }h(x;\lambda )dx\in D(\mathcal{M})$ be an almost everywhere smooth functional on the manifold $\mathcal{M}$, depending parametrically on $\lambda \in \mathbb{C}$, and
whose density satisfies the differential condition%
\begin{gather}
h_{t}=\lambda (uh)_{x}  \label{H1}
\end{gather}%
for all $t\in \mathbb{R}$ and $\lambda \in \mathbb{C}$ on the solution set
of equation \eqref{W5.14}. Then the following iterative differential
relationship
\begin{gather}
(f/h)_{t}=\lambda (uf/h)_{x}  \label{H2}
\end{gather}%
holds, if a smooth function $f\in C^{\infty }(\mathbb{R};\mathbb{R})$
$($parametrically depending on $\lambda \in \mathbb{C})$ satisfies for all $t\in \mathbb{R}$ the linear equation
\begin{gather}
f_{t}=2\lambda u_{x}f+\lambda uf_{x}.  \label{H3}
\end{gather}
\end{proposition}

\begin{proof}
We have from (\ref{H1})--(\ref{H3}) that
\begin{gather*}
(f/h)_{t} = f_{t}/h-fh_{t}/h^{2}=f_{t}/h-\lambda fu_{x}/h-\lambda
fuh_{x}/h^{2}  = f_{t}/h+\lambda fu(1/h)_{x}-\lambda u_{x}f/h \nonumber \\
\phantom{(f/h)_{t}}{}=\lambda (uf)_{x}/h+\lambda uf(1/h)_{x}=\lambda (uf/h)_{x}, 
\end{gather*}
proving the proposition.
\end{proof}

The obvious generalization of the previous proposition is read as follows.

\begin{proposition} \label{Pr_2}If a smooth function $h\in C^{\infty }(\mathbb{R};\mathbb{R})$
satisfies the relationships
\begin{gather*}
h_{t}=ku_{x}h+uh_{x},  
\end{gather*}%
where $k\in \mathbb{R},$ then
\begin{gather*}
H=\int_{0}^{2\pi }h^{1/k}dx  
\end{gather*}%
is a conservation law for the Riemann type hydrodynamical system \eqref{W5.2}.
\end{proposition}

The following polynomial dispersionless functionals, constructed by means of
Proposition~\ref{Pr_2}, are conserved with respect to the f\/low (\ref{W5.15}):
\begin{gather*}
H_{n}^{(1)} :=\int_{0}^{2\pi }dxz^{n}\left(vu_{x}-v_{x}u-\frac{n+2}{n+1}z\right),
 \nonumber\\
H^{(4)} :=\int_{0}^{2\pi
}dx\big[-7v_{x}v^{2}u+z\big(6zu+2v_{x}u^{2}-3v^{2}-4vuu_{x}\big)\big],  \nonumber \\
H^{(5)} :=\int_{0}^{2\pi }dx\big(z^{2}u_{x}-2zvv_{x}\big),\qquad H^{(6)}:=\int_{0}^{2\pi
}dx\big(z_{z}v^{3}+3z^{2}v_{x}u+z^{3}\big),  \nonumber \\
H^{(7)} :=\int_{0}^{2\pi }dx\big(z_{x}v^{3}+3z^{2}vu_{x}-3z^{3}\big),  \nonumber \\
H^{(8)} :=\int_{0}^{2\pi
}dxz\big(6z^{2}u+3zv_{x}u^{2}-3zv^{2}-4zvu_{x}-2v_{x}v^{2}u+2v^{3}u_{x}\big),
\nonumber \\
H^{(9)} :=\int_{0}^{2\pi
}dx\big[1001v_{x}v^{4}u+\big(1092z^{2}u^{2}+364zv_{x}u^{3}-  \nonumber \\
\phantom{H^{(9)} :=}{} -1092zv^{2}u-728zvu_{x}u^{2}-364v_{x}v^{2}u^{2}+273v^{4}+728v^{3}u_{x}u\big)\big],
\nonumber \\
H_{n}^{(2)} :=\int_{0}^{2\pi }dxz_{x}vz^{n},\qquad
H_{n}^{(3)}:=\int_{0}^{2\pi }dxz_{x}\big(v^{2}-2zu\big)^{n},  
\end{gather*}
where $n\in \mathbb{Z}_{+}.$ In particular, as $n=1,2,\dots,$ from (\ref{W5.15}%
) one obtains that%
\begin{gather*}
H_{0}^{(2)}  : =\int_{0}^{2\pi }dxz_{x}v,\qquad H_{1}^{(2)}:=\int_{0}^{2%
\pi }dxz_{x}zv,\qquad \dots,  \nonumber \\
H_{1}^{(3)}  : =\int_{0}^{2\pi }dxz_{x}\big(v^{2}-2uz\big),\qquad
H_{2}^{(3)}  : =\int_{0}^{2\pi }dxz_{x}\big(v^{4}+4z^{2}u^{2}-4zv^{2}u\big),\qquad \dots, 
\end{gather*}
and so on. Similarly one can construct also inf\/inite hierarchies of
conservation laws for the hydrodynamical system~(\ref{W5.15}), which are
both non-polynomial and dispersive:
\begin{gather*}
H_{1}^{(1/4)}  = \int_{0}^{2\pi
}dx\big(-2u_{xx}u_{x}z_{x}+u_{xx}v_{x}^{2}+2u_{x}^{2}z_{xx}   -u_{x}v_{xx}v_{x}+3v_{xx}z_{x}-3v_{x}z_{xx}\big)^{1/4},  \nonumber \\
H_{2}^{(1/3)}  = \int_{0}^{2\pi }dx(-v_{xx}z_{x}+v_{x}z_{xx})^{1/3},\text{ }
\nonumber \\
H_{3}^{(1/3)}  = \int_{0}^{2\pi }dx(v_{xx}u_{x}-v_{x}u_{xx}-z_{xx})^{1/3},
\nonumber \\
H_{1}^{(1/2)}  = \int_{0}^{2\pi
}dx\big[-2vu_{x}z_{x}+v_{x}^{2}+z(-u_{x}v_{x}+3z_{x})\big]^{1/2},  \nonumber \\
H_{2}^{(1/2)}  = \int_{0}^{2\pi
}dx\big(8u_{x}^{3}z_{x}-3u_{x}^{2}v_{x}^{2}-18u_{x}v_{x}z_{x}+6v_{x}^{3}+9z_{x}\big)^{1/2},
 \nonumber\\
H_{1}^{(1/5)}  = \int_{0}^{2\pi
}dx\big(-2u_{xxx}u_{x}z_{x}+u_{xxx}v_{x}^{2}+6u_{xx}^{2}z_{x}-6u_{xx}u_{x}z_{xx}
\nonumber \\
 \phantom{H_{1}^{(1/5)}  =}{}  -3u_{xx}v_{xx}v_{x}+2u_{x}^{2}z_{xxx}-u_{x}v_{xxx}v_{x}
  +3u_{x}v_{xx}^{2}+3v_{xxx}z_{x}-3v_{x}z_{xxx}\big)^{1/5},  \nonumber \\
H_{3}^{(1/3)}  = \int_{0}^{2\pi
}dx\big[k_{1}u(-v_{xx}z_{x}+v_{x}z_{xx})+k_{1}v(u_{xx}z_{x}-u_{x}z_{xx}) 
 \\
 \phantom{H_{3}^{(1/3)}  =}{}  +z(k_{2}u_{xx}v_{x}-k_{2}u_{x}v_{xx}+k_{1}z_{xx}+k_{2}z_{xx})
  +k_{3}(-3u_{x}v_{x}z_{x}+v_{x}^{3}+3z_{x}^{2})\big]^{1/3}, \quad \dots, \nonumber
\end{gather*}
and so on, where $k_{j}\in \mathbb{R}$, $j=1,2,3$, are arbitrary real
numbers. The problem which remains still open consists in proving, if any,
that the generalized hydrodynamical system (\ref{W5.15}) is a Lax type
integrable bi-Hamiltonian f\/low on the periodic functional manifold $\mathcal{%
M}:=C^{(\infty )}(\mathbb{R}/2\pi\mathbb{Z} ;\mathbb{R}^{3}),$ as it was proved above
for the system (\ref{W5.14b}) at $N=2.$ This problem will be analyzed in the
Section below.

\section{The Hamiltonian analysis}

Consider the system (\ref{W5.15}) as a nonlinear dynamical system
\begin{gather}
\left.
\begin{array}{@{}l}
u_{t}=v-uu_{x} \\
v_{t}=z-uv_{x} \\
z_{t}=-uz_{x}
\end{array}
\right\} :=K[u,v,z],  \label{W6.1}
\end{gather}%
on the $2\pi $-periodic smooth functional manifold $\mathcal{M}$ and analyze
it from the Hamiltonian point of view. To tackle with this problem, it is
enough to construct \cite{FT,HPP,PM} exact non-symmetric solutions to the
Lie--Lax equation
\begin{gather}
d\varphi /dt+K^{\prime,\ast }\varphi ={\rm grad}\,L,\qquad
\varphi ^{\prime }\neq \varphi ^{\prime ,\ast },  \label{W6.2}
\end{gather}
for some functional $L\in D(\mathcal{M}),$ where $\varphi \in T^{\ast }(\mathcal{M})$ is, in general, a quasi-local vector, such that the system (\ref{W5.15}) allows the following Hamiltonian representation:
\begin{gather*}
K[u,v,z]=-\eta \,{\rm grad}\, H[u,v,z],\qquad
H=(\varphi ,K)-L,\qquad \eta ^{-1}=\varphi^{\prime }-\varphi
^{\prime ,\ast }.
\end{gather*}
As a test solution to (\ref{W6.2}) one can take the one
\[
\varphi =\big(u_{x}/2,0,-z_{x}^{-1}u_{x}^{2}/2+z_{x}^{-1}v_{x}\big)^{\intercal },
\qquad L=\frac{1}{2}\int_{0}^{2\pi }(2z+vu_{x})dx,
\]
which gives rise to the following co-implectic operator:
\begin{gather}
\eta ^{-1}:=\varphi ^{\prime }-\varphi ^{\prime ,\ast }=\left(
\begin{array}{ccc}
\partial & 0 & -\partial u_{x}z_{x}^{-1} \vspace{1mm}\\
0 & 0 & \partial z_{x} \vspace{1mm}\\
-u_{x}z_{x}^{-1}\partial & z_{x}\partial &
\begin{array}{c}
\frac{1}{2}(u_{x}^{2}z_{x}^{-2}\partial +\partial u_{x}^{2}z_{x}^{-2}) \\
{} -(v_{x}z_{x}^{-2}\partial +\partial v_{x}z_{x}^{-2})
\end{array}
\end{array}
\right) .  \label{W6.4}
\end{gather}%
This expression is not strictly invertible, as its kernel possesses the
translation vector f\/ield $d/dx:\mathcal{M}\rightarrow T(\mathcal{M})$ with
components $(u_{x},v_{x},z_{x})^{\intercal }\in T(\mathcal{M}),$ that is $%
\eta ^{-1}(u_{x},v_{x},z_{x})^{\intercal }=0.$

Nonetheless, upon formal inverting the operator expression (\ref{W6.4}), we
obtain by means of simple enough, but slightly cumbersome, direct
calculations, that the Hamiltonian function equals
\begin{gather}
H:=\int_{0}^{2\pi }dx(u_{x}v-z).  \label{W6.5}
\end{gather}%
and the implectic $\eta $-operator looks as
\begin{gather}
\eta :=\left(
\begin{array}{ccc}
\partial ^{-1} & u_{x}\partial ^{-1} & 0 \vspace{1mm}\\
\partial ^{-1}u_{x} & v_{x}\partial ^{-1}+\partial ^{-1}v_{x} & \partial
^{-1}z_{x} \vspace{1mm}\\
0 & z_{x}\partial ^{-1} & 0
\end{array}
\right) .  \label{W6.6}
\end{gather}

The same way, representing the Hamiltonian function (\ref{W6.5}) in the
scalar form%
\begin{gather}
H=(\psi ,(u_{x},v_{x},z_{x})^{\intercal }),\qquad \psi =\frac{1}{2}\left(-v,u+\cdots ,-\frac{1}{\sqrt{z}}\partial ^{-1}\sqrt{z}\right)^{\intercal },
\label{W6.7}
\end{gather}
one can construct a second implectic (co-symplectic) operator $\vartheta
:T^{\ast }(\mathcal{M})\rightarrow T(\mathcal{M})$, looking up to $O(\mu
^{2})$ terms, as follows:
\begin{gather}
\vartheta =\left(\!
\begin{array}{@{}c@{\,\,\,}c@{\,\,\,}c@{}}
\mu \left(\frac{(u^{(1)})^{2}}{z^{(1)}}\partial +\partial \frac{(u^{(1)})^{2}}{%
z^{(1)}}\right)
&
\begin{array}{@{}c@{}}
1+\frac{2\mu }{3}\left( \frac{u^{(1)}v^{(1)}}{z^{(1)}}\partial\right.\\
\left.{} +2\partial \frac{%
u^{(1)}v^{(1)}}{z^{(1)}}\right)
\end{array}
&
\frac{2\mu }{3}\left(\partial \frac{(v^{(1)})^{2}}{z^{(1)}}+\partial u^{(1)}\right)
\vspace{2mm}\\
\begin{array}{@{}c@{}}
-1+\frac{2\mu }{3}\left(\partial \frac{u^{(1)}v^{(1)}}{z^{(1)}}\right.\\
\left.{}+2\frac{%
u^{(1)}v^{(1)}}{z^{(1)}}\partial \right)
\end{array}&
\begin{array}{@{}c@{}}
\frac{2\mu }{3}\left(\frac{(v^{(1)})^{2}}{z^{(1)}}\partial +\partial \frac{%
(v^{(1)})^{2}}{z^{(1)}}\right){} \\
{}+\frac{2\mu }{3}\big(u^{(1)}\partial +\partial u^{(1)}\big)
\end{array}
& 2\mu \partial v^{(1)} \vspace{2mm}\\
\frac{2\mu }{3}\left(\frac{(v^{(1)})^{2}}{z^{(1)}}\partial +u^{(1)}\partial \right)%
&
2\mu v^{(1)}\partial%
&
\mu \big(\partial z^{(1)}+z^{(1)}\partial \big)%
\end{array}\!
\right)\! +O(\mu ^{2}),\!\!  \label{W6.8}
\end{gather}%
where we put, by def\/inition, $\vartheta ^{-1}:=(\psi ^{\prime }-\psi
^{\prime ,\ast })$, $u:=\mu u^{(1)}$, $v:=\mu v^{(1)}$, $z:=\mu z^{(1)}$ as $\mu
\rightarrow 0,$ and whose exact form needs some additional simple enough but
cumbersome calculations, which will be presented in a work under preparation.

The operator (\ref{W6.8}) satisf\/ies the Hamiltonian vector f\/ield condition:
\begin{gather*}
(u_{x},v_{x},z_{x})^{\intercal }=-\vartheta \, {\rm grad}\, H,
\end{gather*}%
following easily from~(\ref{W6.7}).

Now having applied to the pair of implectic operators the gradient-holonomic
scheme \cite{HPP,MBPS,PM} of constructing a Lax type representation for the
dynamical system (\ref{W6.1}) we obtain by means of slightly cumbersome and
tedious calculations the following compatible Lax type representation:
\begin{gather}
f_{x}=\ell \lbrack u,v;\lambda ]f,\qquad f_{t}=p(\ell )f,\qquad
p(\ell ):=-u\ell \lbrack u,v;\lambda ]+q(\lambda ), \nonumber\\
\ell \lbrack u,v,z;\lambda =\left(
\begin{array}{ccc}
\lambda u_{x} & -v_{x} & z_{x} \\
3\lambda ^{2} & -2\lambda u_{x} & \lambda v_{x} \\
\lambda ^{2}r[u,v,z] & -3\lambda & \lambda u_{x}%
\end{array}%
\right) ,\qquad q(\lambda ):=\left(
\begin{array}{ccc}
0 & 0 & 0 \\
\lambda & 0 & 0 \\
0 & 1 & 0%
\end{array}%
\right) , \nonumber
\\
 p(\ell )=\left(
\begin{array}{ccc}
-\lambda uu_{x} & uv_{x} & -uz_{x} \\
-3u\lambda ^{2}+\lambda & 2\lambda uu_{x} & -\lambda uv_{x} \\
-\lambda ^{2}r[u,v,z]u & 1+3u\lambda & -\lambda uu_{x}%
\end{array}%
\right) ,
\label{W6.10}
\end{gather}%
where $f\in C^{\infty }(\mathbb{R};\mathbb{C}^{3})$,  $\lambda \in \mathbb{C} \backslash \{0\}$ is a spectral parameter and $r:\mathcal{M}\rightarrow
\mathbb{R}$ is a smooth mapping, satisfying the dif\/ferential equation
\begin{gather*}
D_{t}r+u_{x}r=6.  
\end{gather*}
The latter possesses a wide set $\mathcal{R}$ of dif\/ferent solutions,
amongst which there are the following:%
\begin{gather}
r  \in  \mathcal{R}:=\Bigg\{\big[\big(6xv-3u^2\big)/z\big]_{x},\; 3\big(2v_{x}-u_{x}^{2}\big)z_{x}^{-1},\;
\frac{2u_{x}^{3}-6u_{x}v_{x}+9z_{x}}{2u_{x}z_{x}-v_{x}^{2}}, \nonumber
\\
 \phantom{r  \in  \mathcal{R}:=\Bigg\{}{} \big(v_{x}v^{3}-3u_{x}v^{2}z+uz_{x}\big(uz-v^{2}\big)+6vz^{2}\big)z^{-3}\Bigg\}.  \label{W6.12}
\end{gather}%
Thereby, the following proposition holds.

\begin{proposition}
The generalized Riemann type hydrodynamical equation \eqref{W5.14b} at $N=2$
and $N=3$ is equivalent to Lax type integrable bi-Hamiltonian dynamical
systems \eqref{W5.2} and \eqref{W6.1}, whose Hamiltonian structures and Lax
type representations are given by expressions  \eqref{W5.7-3}, \eqref{W5.10},  \eqref{W5.3},   and \eqref{W6.6}, \eqref{W6.8},  \eqref{W6.10}, \eqref{W6.12}, respectively.
\end{proposition}

Note here that only the third element from the set \eqref{W6.12} allows the
reduction $z=0$ to the case $N=2.$   Concerning the case $N=4$ and the general case $N\in
\mathbb{Z}_{+}$,   applying successively the method devised above, one can
obtain~\cite{BGPPP} for the Riemann type hydrodynamical system~\eqref{W6.1} both inf\/inite
hierarchies of   dispersive and dispersionless conservation laws, their
symplectic structures and the related Lax type representations, which is a
topic of the next work under preparation.

\subsection*{Acknowledgments}

M.P. and A.P. are appreciated to Organizers of the
Symmetry-2009 Conference (June 21--27, 2009) held in Kyiv, Ukraine, and
the NEEDS-2009 Conference (May 15--23,  2009), held in Isola Rossa of Sardinia,
Italy, for the invitations to deliver reports and for a partial support.
M.P. was, in part, supported by RFBR grant 08-01-00054 and a grant of the
RAS Presi\-dium ``Fundamental Problems of Nonlinear Dynamics''. The authors
thanks go to Professors M.~B\l aszak,  N.~Bogolubov (jr.) and D.~Blackmore for useful discussions of the results obtained. Authors are also
cordially thankful to Referees who have read the article and made very
important remarks  and suggestions, which were very instrumental for f\/inal
preparing a manuscript, and which made it possible both to improve and
correct the exposition.

\pdfbookmark[1]{References}{ref}
\LastPageEnding

\end{document}